\documentclass[a4paper,12pt]{article}
\usepackage{cite}
\usepackage{amssymb}
\usepackage{amsmath}
\usepackage{amsfonts}
\usepackage{amsthm}
\usepackage{mathtools}
\usepackage[utf8]{inputenc}
\usepackage{color}
\usepackage{graphicx}
\usepackage{color}
\usepackage[active]{srcltx}

\usepackage{here}
\usepackage{cite}
\usepackage{soul}
\usepackage[affil-it]{authblk}

\usepackage{multicol}

\usepackage{titlesec}
\titleformat{\subsection}
       {\normalfont\fontsize{14}{17}\itshape}{\thesubsection}{1em}{}

\input paperdef

\begin{document}

\null\hfill IFT-UAM/CSIC-24-159

\thispagestyle{empty}

\title{
From the Unification of Conformal and Fuzzy Gravities with Internal
Interactions to the $SO(10)$ GUT and the Particle Physics Standard Model}
\date{}
\author{Gregory Patellis$^{1}$\thanks{email: grigorios.patellis@tecnico.ulisboa.pt}~, 
Nicholas Tracas$^2$\thanks{email: ntrac@central.ntua.gr}~
and George Zoupanos$^{2,3,4,5}$\thanks{email: george.zoupanos@cern.ch}\\
{\small
$^1$ Centro de Física Teórica de Partículas - CFTP, Departamento de Física,\\ Instituto Superior Técnico, Universidade de Lisboa,\\
Avenida Rovisco Pais 1, 1049-001 Lisboa, Portugal \\
$^2$ Physics Department,   National Technical University, 157 80 Zografou, Athens, Greece\\
$^3$ Instituto de Física Teórica, (UAM/CSIC), Universidad Autónoma de Madrid, Cantoblanco, 28049, Spain\\
$^4$ Max-Planck Institut f\"ur Physik, Boltzmannstr. 8, 85748 Garching, Germany,\\
$^5$ Institut für Theoretische Physik der Universität Heidelberg,
Philosophenweg 16, 69120 Heidelberg, Germany \\
}
}

{\let\newpage\relax\maketitle}

\begin{abstract}
\noindent In the present study, the unification of the Conformal and Fuzzy Gravities with the Internal
Interactions is based on the observation that the tangent space of a
curved space and the space itself do not have necessarily the same
dimensions. Moreover, the construction is based on the fact that the
gravitational theories can be formulated in a gauge-theoretical way.
In the present work we study the various consecutive breakings through which these unified theories can ultimately result into the Standard Model. We estimate the scales of the breakings in each case using one-loop RGEs.
\end{abstract}


\section{Introduction}\label{intro}

The target of constructing a unified scheme involving all 
interactions was in the center of discussions in the physics 
community since many decades and started more than a century ago 
with the early attempts of Kaluza and Klein \cite{Kaluza:1921, Klein:1926}, who were trying 
to unify the known interactions at the time, namely gravity and 
electromagnetism, by going to five dimensions. An interesting 
revival of the Kaluza-Klein scheme started much later when it was 
realised that non-abelian gauge groups, such as those that 
constitute the well established today Standard Model (SM) of 
Particle Physics, appear too naturally in addition to the $U(1)$ of
electromagnetism when one considers further extensions of the space 
dimensions \cite{Kerner:1968,CHO1987358,PhysRevD.12.1711}. Assuming that the total space-time can be described as a direct product structure 
$M_D=M_4 \times B$, 
where $M_4$ is the usual Minkowski space-time and $B$ is a compact 
Riemannian space with a non-abelian isometry
group $S$, the dimensional reduction of the theory leads to gravity 
coupled to a Yang-Mills theory with a gauge group containing $S$ and 
scalars in four dimensions. 
Therefore an attractive geometric unification of gravity with other 
interactions, potentially those of the SM, is achieved together with 
an explanation of the gauge symmetries. However there exist serious 
problems in the Kaluza-Klein framework, such as that there is no
classical ground state corresponding to the direct product structure 
of $M_D$. From the Particle Physics point of view though the most 
serious problem is that, after adding fermions to the original 
action, 
it is impossible to obtain chiral fermions in four dimensions \cite{Witten:1983}.
Eventually this problem is resolved by adding suitable matter fields 
to the original gravity action in particular Yang-Mills at the cost 
of abandoning the pure geometric unification. Accepting the fact 
that one has to introduce Yang-Mills fields in higher dimensions and 
considering that they are part of a Grand Unified Theory (GUT) 
together with a Dirac one \cite{Georgi1999,FRITZSCH1975193}, the restriction to obtain chiral 
fermions in four dimensions is limited in requiring that the total 
dimension of spacetime has to be of the form $4k + 2$ (see e.g., 
ref. \cite{CHAPLINE1982461}). 

During the last several decades the Superstring theories 
(see e.g., refs. \cite{Green2012-ul,polchinski_1998,Lust:1989tj}) dominated the research on extra 
dimensions. In particular the heterotic string theory \cite{GROSS1985253}, defined 
in ten dimensions, was the most promising, due to the fact that the 
SM gauge group can be accommodated into GUTs that emerge after the 
dimensional
reduction of the initial $E_8 \times E_8$ of the theory. It should 
be 
noted though that even before the formulation of superstring 
theories, another framework has been developed that focused on the 
dimensional reduction of
higher-dimensional gauge theories, which provided another avenue
for exploring the unification of fundamental interactions \cite{forgacs,MANTON1981502,Kubyshin:1989vd,KAPETANAKIS19924,LUST1985309}.
The latter approach to unify fundamental interactions, which shared
common objectives with the superstring theories, was first 
investigated by
Forgacs-Manton (F-M) and Scherk-Schwartz (S-S). F-M explored the 
concept of Coset Space Dimensional Reduction (CSDR)\cite{forgacs,MANTON1981502,Kubyshin:1989vd,KAPETANAKIS19924}, which can 
lead naturally to chiral fermions while S-S focused on the group 
manifold reduction \cite{SCHERK197961}, which does not admit chiral fermions 
though. Recent developments and attempts towards realistic models that can be 
confronted with experiment can be found in refs. \cite{Manousselis_2004,Chatzistavrakidis:2009mh,Irges:2011de,Manolakos:2020cco,Patellis:2023hep}.

Given the above picture concerning the gauge theories part of 
unification attempts on the gravity side diffeomorphism-invariant 
gravity theory is obviously invariant with respect to 
transformations whose parameters are
functions of spacetime, just as in the local gauge theories. 
Therefore, naturally it has been long believed that general 
relativity (GR) can be
formulated as a gauge theory \cite{PhysRev.101.1597,Kibble:1961ba,PhysRevLett.38.739,Ivanov:1980tw,Ivanov:1981wn,PhysRevD.21.1466,Kibble:1985sn}. 
The gauge-theoretical approach 
to gravity started with Utiyama's pioneering study \cite{PhysRev.101.1597}.
The spin connection can be treated as the gauge field of 
the theory and would enter in the action using its field strength. 
Subsequently more elegantly in addition to the spin connection the 
veilbein was identified also as part of the gauge fields and it was 
discussed as a gauge theory of the de Sitter $SO(1,4)$ or the AdS 
group $SO(2,3)$, spontaneously broken by a scalar field to the 
Lorentz $SO(1,3)$ group \cite{PhysRevD.21.1466,Kibble:1985sn}.  In addition, the 
Conformal Gravity (CG) has been formulated by gauging the conformal group $SO(2,4)$ \cite{Kaku:1978nz,Roumelioti:2024lvn}, which can be spontaneously broken to the well-known Eistein Gravity (EG) or 
to the Weyl gravity \cite{Roumelioti:2024lvn}, a model whose action is made out of the 
square of the Weyl tensor. The latter has been studied quite 
extensively \cite{Fradkin:1985am,Maldacena:2011mk,Mannheim:2011ds,Anastasiou:2016jix,Ghilencea:2018dqd,Hell:2023rbf,Ghilencea:2023sti}. The basic 
idea of the gauge-theoretic approach to gravity was used in a 
fundamental way in supergravity (see e.g. ref. \cite{freedman_vanproeyen_2012,Ortin:2015hya,Castellani:2013iq}) while 
recently it was transferred in the non-commutative gravity attempts 
too \cite{Chatzistavrakidis_2018,Manolakos:2019fle,Manolakos:2021rcl,Manolakos:2022ihc,Manolakos:2023hif,Roumelioti:2024jib}.

A more interesting suggestion towards unifying gravity as gauge 
theory with the rest fundamental interactions described as GUTs was 
suggested in the past \cite{Percacci:1984ai,Percacci_1991} and revived recently \cite{Nesti_2008,Nesti_2010,Krasnov:2017epi,Chamseddine2010,Chamseddine2016,noncomtomos,Konitopoulos:2023wst,Roumelioti:2024lvn,Manolakos:2023hif}.
It is based on the following observation: although usually the 
dimension of the tangent space is taken to be equal to the dimension 
of the corresponding curved manifold, the tangent group of a manifold of dimension $d$ is not necessarily $SO(d)$ \cite{Weinberg:1984ke}. It is 
very interesting that one can consider higher than four dimensional 
tangent groups in a four dimensional space-time which opens the 
possibility to achieve unification of gravity with internal 
interactions by gauging these higher-dimensional tangent groups. 
Technically a very attractive feature of this approach is that most 
of the machinery that was used in the high dimensional theories with 
extra physical space dimensions, such as in the coset space 
dimensional reduction (CSDR) scheme \cite{CHAPLINE1982461,forgacs,KAPETANAKIS19924,Kubyshin:1989vd,SCHERK197961,MANTON1981502,LUST1985309,Manousselis_2004,Chatzistavrakidis:2009mh,Irges:2011de,Manolakos:2020cco,Patellis:2023hep}, can be transfered in 
the present four-dimensional constructions since they have the same 
tangent group. For instance there exist constraints which one has to 
take into account when aiming to construct realistic chiral theories 
for the internal interactions; similarly attempting to avoid 
a doubling of the spectrum of chiral theories by imposing Majorana 
condition in addition to Weyl in the extra dimensions \cite{CHAPLINE1982461,KAPETANAKIS19924}.

Along the lines described above, a unification of CG and internal interactions has been constructed recently \cite{Roumelioti:2024lvn}. 
This construction was subsequently extended to the unification of 
four-dimensional Gravity on a covariant noncommutative (fuzzy) space 
(fuzzy gravity - FG)  with internal interactions \cite{Roumelioti:2024jib}. In the present work we 
focus on the consequences of the  
 CG - internal interactions unification scheme w.r.t the physics that follow in lower energy scales, while we also comment on the  FG - internal interactions unification case, since by construction they have several similarities.

\section{Conformal Gravity}\label{CG}

According to the discussion in \refse{intro}, EG has been treated as the gauge theory  of the Poincar\'e 
group but much insight and elegance was gained by considering 
instead the gauging of the de Sitter (dS), $SO(1,4)$, and the 
Anti-de Sitter group (AdS), $SO(2,3)$. Both of these groups contain 
the same number of generators, i.e. $10$ as the Poincar\'e group and can 
be spontaneously broken by a scalar field to the Lorentz group, 
$SO(1,3)$ \cite{PhysRevD.21.1466,Kibble:1985sn, manolakosphd, Roumelioti:2024lvn}. The Poincar\'e, the 
dS and the AdS groups are all subgroups of the conformal group 
$SO(2,4)$, which has 15 generators and is the group of 
transformations on space-time which leave invariant the null 
interval $ds^2 = \eta_{\mu\nu} dx^\mu dx^\nu = 0$.
In ref \cite{Kaku:1978nz} the gauge theory formalism of Gravity was 
extended to the conformal group constructing in this way the CG. The breaking of CG to EG or to Weyl’s scale invariant theory of
gravity was done by the imposition of constraints (see e.g. \cite{Kaku:1978nz}). 
However in ref \cite{Roumelioti:2024lvn}, for the first time, the breaking of the 
conformal gauge group was done spontaneously, induced by the 
introduction of a scalar field in the action using the Lagrange 
multiplier method.

\subsection{Spontaneous symmetry breaking}

The spontaneous symmetry breaking of CG, which is based, as already 
mentioned on the $SO(2,4)$ gauge group, whose algebra is isomorphic 
to those of $SU(4)$ and $SO(6)$, can be done in two ways. For 
convenience we work with Euclidean signature. Then one way is to 
introduce a scalar in the vector representation (rep) of $SO(6)$, $6$, 
which takes vev in the $\langle 1\rangle$ component \cite{Slansky:1981yr,Feger:2019tvk} of the 6 
according to the branching rules of reps of $SO(6)$ to its maximal 
subgroup $SO(5)$ \cite{Roumelioti:2024lvn}, i.e.
\begin{equation}
\begin{aligned} 
    SO(6) &\supset SO(5)\\
    6 &=1+5
\end{aligned}
\end{equation}
Then the $SO(5)$, being isomorphic to $SO(2,3)$, can break 
spontaneously further to $SO(1,3)$, when a scalar in the $5$ rep takes 
a vev in the $\langle 1,1\rangle$ component according to the 
branching rules:
\begin{equation}
\begin{aligned} 
    SO(5) &\supset SU(2)\times SU(2)\\
    5 &=(1,1)+(2,2),
\end{aligned}
\end{equation}
where the algebra of $SU(2)\times SU(2)$ is isomorphic to those of 
$SO(4)$ 
and $SO(1,3)$. Therefore in order to realise the above breakings 
from $SO(2,4)$ to $SO(1,3)$ one has to introduce two scalar fields, 
belonging in the vector rep, $6$ of $SO(2,4)$ (for details see ref 
\cite{Roumelioti:2024lvn}).

Keeping in mind the above way of breaking the SO(2,4) to SO(1,3), we can use a more direct way to achieve the same breaking in one step using a scalar belonging in the 2nd rank antisymmetric rep, 15.
In fact this breaking can lead either to the four-dimensional EG or to WG as we will see
\footnote{The same breaking can be achieved in the two steps breaking too (see ref \cite{Roumelioti:2024lvn}).}.

The gauge group $SO(2, 4)$, as mentioned previously, has 15 
generators. These generators in four-dimensional  notation can be 
represented by the six Lorentz transformations, $M_{ab}$, four translations, $P_a$ , four special conformal transformations (conformal boosts), $K_a$, and the dilatation
\footnote{The details 
on the reps chosen, along with the commutation and anticommutation 
relations of the generators, can be found in ref \cite{Roumelioti:2024lvn}.}.

\subsection{Einstein-Hilbert and Weyl action from SSB of the conformal group, by using a scalar in the adjoint representation}
\label{AdjointSubsection}

In the past, in order to construct the four-dimensional CG, one had to start by gauging the conformal group, $SO(2,4)$, and impose constraints in order to retrieve WG (see e.g. \cite{Kaku:1978nz}). Here, instead, we use SSB mechanism which
is more elegant and appropriate for a field theory treatment, and the $SO(2,4)$ gauge group is being directly reduced to the $SO(1,3)$ by a scalar field belonging to the adjoint rep. This can lead either to the four-dimensional EG or to WG.

The gauge group, $SO(2,4)$, as mentioned in the previous subsection, comprises of fifteen generators. Those generators in four-dimensional notation consist of six Lorentz transformations, $M_{ab}$, four translations, $P_a$, four special conformal transformations (conformal boosts), $K_a$, and the dilatation, $D$.

The gauge connection, $A_\mu$, as an element of the $SO(2,4)$ algebra, can be expanded in terms of the generators as
\begin{equation}
A_\mu= \frac{1}{2}\omega_\mu{}^{a b} M_{a b}+e_\mu{}^a P_a+b_\mu{}^a K_a+\tilde{a}_\mu D,
\end{equation}
where, for each generator a gauge field has been introduced. The gauge field related to the translations is identified as the vierbein, while the one of the Lorentz transformations is identified as the spin connection. The field strength tensor is of the form
\begin{equation}\label{fst}
F_{\mu \nu}=\frac{1}{2}R_{\mu \nu}{}^{a b} M_{a b}+\tilde{R}_{\mu \nu}{}^a P_a+R_{\mu \nu}{}^a K_a+R_{\mu \nu} D,
\end{equation}
where
\begin{equation}\label{curves}
\begin{aligned}
R_{\mu \nu}{}^{a b} & =\partial_\mu \omega_\nu{}^{a b}-\partial_\nu \omega_\mu{}^{a b}-\omega_\mu{}^{a c} \omega_{\nu c}{}^b+\omega_\nu{}^{a c} \omega_{\mu c}{}^b-8 e_{[\mu}{}^{[a} b_{\nu]}{}^{b]} \\
& =R_{\mu \nu}^{(0) a b}-8 e_{[\mu}{}^{[a} b_{\nu]}{}^{b]}, \\
\tilde{R}_{\mu \nu}{}^a & =\partial_\mu e_\nu{}^a-\partial_\nu e_\mu{}^a+\omega_\mu{}^{a b} e_{\nu b}-\omega_\nu{}^{a b} e_{\mu b}-2 \tilde{a}_{[\mu} e_{\nu]}{}^a \\
& =T_{\mu \nu}^{(0) a}-2 \tilde{a}_{[\mu} e_{\nu]}{}^a, \\
R_{\mu \nu}{}^a & =\partial_\mu b_\nu{}^a-\partial_\nu b_\mu{}^a+\omega_\mu{}^{a b} b_{\nu b}-\omega_\nu{}^{a b} b_{\mu b}+2 \tilde{a}_{[\mu} b_{\nu]}{}^a\\
&=T_{\mu \nu}^{(0) a}(b)+2 \tilde{a}_{[\mu} b_{\nu]}{}^a,\\
R_{\mu \nu} & =\partial_\mu \tilde{a}_\nu-\partial_\nu \tilde{a}_\mu+4 e_{[\mu}{}^a b_{\nu] a},
\end{aligned}
\end{equation}
where $T_{\mu \nu}^{(0) a}$ and $R_{\mu \nu}^{(0) a b}$ are the torsion and curvature component tensors in the four-dimensional vierbein formalism of GR, while $T_{\mu \nu}^{(0) a}(b)$ is the torsion tensor related to the gauge field $b_\mu{}^a$. 

We shall start by choosing the parity conserving action, which is quadratic in terms of the field strength tensor \eqref{fst}, in which we have introduced a scalar that belongs in the 2nd rank antisymmetric rep, $15$, of $SO(6) \sim SO(2,4)$ along with a dimensionful parameter, $m$:
\begin{equation}
    S_{SO(2,4)}=a_{CG}\int d^4x [\operatorname{tr} \epsilon^{\mu \nu \rho \sigma} m\phi F_{\mu \nu}F_{\rho \sigma}+(\phi^2-m^{-2} \mathbb{I}_4)], 
\end{equation}
where the trace is defined as $ \operatorname{tr}\rightarrow \epsilon_{abcd} [\text{Generators}]^{abcd}$.

The scalar expanded on the generators is:
\begin{equation}
\phi=\phi^{a b} M_{a b}+\tilde{\phi}^a P_a+\phi^a K_a+\tilde{\phi} D~.
\end{equation}

In accordance with \cite{Li:1973mq}, we pick the specific gauge in which $\phi$ is diagonal of the form $\operatorname{diag}(1,1,-1,-1)$. Specifically we choose $\phi$ to be only in the direction of the dilatation generator $D$:
\begin{equation}
    \phi=\phi^0=\tilde{\phi}D \xrightarrow{\phi^2=m^{-2}\mathbb{I}_4}\phi=-2m^{-1} D.
\end{equation}
In this particular gauge the action reduces to
\begin{equation}
    S=-2a_{CG}\int d^4x \operatorname{tr} \epsilon^{\mu \nu \rho \sigma} F_{\mu \nu}F_{\rho \sigma}D, 
\end{equation}
and the gauge fields $e,b$ and $\tilde{a}$ become scaled as $me,mb$ and $m\tilde{a}$ correspondingly.
After straightforward calculations, using the expansion of the field strength tensor as in eq. \eqref{fst}, and the anticommutation relations of the generators, we obtain: 
\begin{equation}
\begin{gathered}
    S=-2a_{CG}\int d^4x \operatorname{tr} \epsilon^{\mu \nu \rho \sigma}\Big[\frac{1}{4}R_{\mu \nu}{}^{ab}R_{\rho \sigma}{}^{cd}M_{ab}M_{cd}D\\
    +i\epsilon_{abcd}(R_{\mu \nu}{}^{ab}R_{\rho \sigma}{}^{c} K^d D - R_{\mu \nu}{}^{ab}\tilde{R}_{\rho \sigma}{}^{c}P^{d}D)+(\frac{1}{2}\tilde{R}_{\mu \nu}{}^{a}R_{\rho \sigma} + 2\tilde{R}_{\mu \nu}{}^{a}R_{\rho \sigma}{}^{b})M_{ab}\\
    +(\frac{1}{4}R_{\mu \nu}R_{\rho \sigma}- 2\tilde{R}_{\mu \nu}{}^{a}R_{\rho \sigma a})D   
    \Big].
\end{gathered}
\end{equation}
In this point we employ the trace on the several generators and their products. In particular:
\begin{equation}
\begin{gathered}
   \operatorname{tr}[K^{d}D]=\operatorname{tr}[P^{d}D]=\operatorname{tr}[M_{ab}]= \operatorname{tr}[D]=0, \\
        \text{and}\quad \operatorname{tr}[M_{ab}M_{cd}D]=-\frac{1}{2}\epsilon_{abcd}.
\end{gathered}
\end{equation}
The resulting broken action is:
\begin{equation}
\label{BrokenActionConformal}
     S_{\mathrm{SO}(1,3)}=\frac{a_{CG}}{4}\int d^4x \epsilon^{\mu \nu \rho \sigma}\epsilon_{abcd}R_{\mu \nu}{}^{ab}R_{\rho \sigma}{}^{cd},
\end{equation}
while its invariance has obviously been reduced only to Lorentz. Before continuing, we notice that there is no term containing the field $\tilde{a}_\mu$ in any way present in the action. Thus, we may set $\tilde{a}_\mu=0$\footnote{ Let us note that since $\tilde{a}_\mu$ is the gauge field corresponding to the dilatation generator, $D$, by switching off $\tilde{a}_\mu$ or by making it heavy due to SSB, it remains a global symmetry corresponding to scale invariance. The latter is broken by the presence of dimensionful parameters as the cosmological constant.\label{footnote}}. This simplifies the form of the two component field strength tensors related to the $P$ and $K$ generators:
\begin{equation}
\begin{aligned}
  \tilde{R}_{\mu \nu}{}^a & =mT_{\mu \nu}^{(0) a}-2 m^2\tilde{a}_{[\mu} e_{\nu]}{}^a \longrightarrow mT_{\mu \nu}^{(0) a}, \\
R_{\mu \nu}{}^a &=mT_{\mu \nu}^{(0) a}(b)+2m^2 \tilde{a}_{[\mu} b_{\nu]}{}^a \longrightarrow mT_{\mu \nu}^{(0) a}(b).
\end{aligned}
\end{equation}
The absence of the above field strength tensors in the action, allows us to also set $\tilde{R}_{\mu \nu}{}^a=R_{\mu \nu}{}^a=0$, and thus to obtain a torsion-free theory. Since $R_{\mu \nu}$ is also absent from the expression of the broken action, it may also be set equal to zero. From its definition in eq. \eqref{curves}, then we obtain the following relation among $e$ and $b$: 
\begin{equation}\label{ef}
    e_\mu{}^a b_{\nu a}-e_{\nu}{}^{a}b_{\mu a}=0,
\end{equation}
The above result reinforces one to consider solutions that relate $e$ and $b$. Here we examine two possible solutions of eq. \eqref{ef}.

\subsubsection{When $b_\mu{}^a=ae_\mu{}^a$ - Einstein-Hilbert action in the presence of a cosmological constant}
\label{subsubsectionA}
In this case, first proposed in \cite{Chamseddine:2002fd}, by a simple substitution we obtain:
\begin{equation}
\begin{aligned}
        S_{\mathrm{SO}(1,3)} =\frac{a_{CG}}{4}\int d^4 x \epsilon^{\mu \nu \rho \sigma} \epsilon_{a b c d}&\left[R_{\mu \nu}^{(0) a b}-4m^2a\left(e_\mu{}^a e_\nu{}^b-e_\mu{}^b e_\nu{}^a\right)\right]\\
       &\left[R_{\rho \sigma}^{(0) c d}-4m^2a\left(e_\rho{}^c e_\sigma{}^d-e_\rho{}^d e_\sigma{}^c\right)\right]\longrightarrow\\
        S_{\mathrm{SO}(1,3)} =\frac{a_{CG}}{4}\int d^4 x \epsilon^{\mu \nu \rho \sigma} \epsilon_{a b c d}&[R_{\mu \nu}^{(0) a b}-8m^2ae_\mu{}^a e_\nu{}^b]\left[R_{\rho \sigma}^{(0) c d}-8m^2ae_\rho{}^c e_\sigma{}^d\right],
       \end{aligned}
\end{equation}
which yields
\begin{equation}\label{so24finalaction}
\begin{aligned}
             S_{\mathrm{SO}(1,3)}=\frac{a_{CG}}{4}\int d^4 x \epsilon^{\mu \nu \rho \sigma} \epsilon_{a b c d}[R_{\mu \nu}^{(0) a b}R_{\rho \sigma}^{(0) c d}-16m^2aR_{\mu \nu}^{(0) a b}e_\rho{}^c e_\sigma{}^d+\\
             +64m^4a^2 e_\mu{}^a e_\nu{}^b e_\rho{}^c e_\sigma{}^d].
\end{aligned}
\end{equation}

This action consists of three terms: one G-B topological term, the E-H action, and a cosmological constant, and for $a<0$ describes EG in AdS space.

\subsubsection{When $b_\mu{}^a=-\frac{1}{4}(R_\mu{}^a+\frac{1}{6}R e_\mu{}^a)$ - Weyl action}

This relation among $b$ and $e$, which is again solution of \eqref{ef}, was suggested in refs \cite{Kaku:1978nz} and \cite{freedman_vanproeyen_2012}. Taking this into account we obtain the following action:
\begin{equation}
    \begin{aligned}
             S=\frac{a_{CG}}{4}\int d^4 x \epsilon^{\mu \nu \rho \sigma} \epsilon_{a b c d}&\left[R_{\mu \nu}^{(0) a b}+\frac{1}{2}\left(m e_\mu{}^{[a} R_\nu{}^{b]}-me_\nu{}^{[a} R_\mu{}^{b]}\right)-\frac{1}{3} m^2 R e_\mu{}^{[a} e_\nu{}^{b]}\right]\\
             &\left[R_{\rho \sigma}^{(0) c d}+\frac{1}{2}\left(m e_\rho{}^{[c} R_\sigma{}^{d]}-m e_\sigma{}^{[c} R_\rho{}^{d]}\right)-\frac{1}{3} m^2 R e_\rho{}^{[c} e_\sigma{}^{d]}\right].
    \end{aligned}
\end{equation}
Considering the rescaled vierbein $\tilde{e}_\mu{}^{a}=m e_\mu{}^{a}$ and recalling that $R_{\mu \nu}^{(0) a b}=-R_{\nu \mu}^{(0) a b}$, we obtain
\begin{equation}
    \begin{aligned}
             S=\frac{a_{CG}}{4}\int d^4 x \epsilon^{\mu \nu \rho \sigma} \epsilon_{a b c d}&\left[R_{\mu \nu}^{(0) a b}-\frac{1}{2}\left(\tilde{e}_\mu{}^{[a} R_\nu{}^{b]}-\tilde{e}_\nu{}^{[a} R_\mu{}^{b]}\right)+\frac{1}{3} R \tilde{e}_\mu{}^{[a} \tilde{e}_\nu{}^{b]}\right]\\
             &\left[R_{\rho \sigma}^{(0) c d}-\frac{1}{2}\left( \tilde{e}_\rho{}^{[c} R_\sigma{}^{d]}- \tilde{e}_\sigma{}^{[c} R_\rho{}^{d]}\right)+\frac{1}{3} R \tilde{e}_\rho{}^{[c} \tilde{e}_\sigma{}^{d]}\right],
    \end{aligned}
\end{equation}
which is equal to 
\begin{align}
             S=\frac{a_{CG}}{4}\int d^4 x \epsilon^{\mu \nu \rho \sigma} \epsilon_{a b c d}C_{\mu \nu}{}^{a b}C_{\rho \sigma}{}^{c d},
\end{align}
where $C_{\mu \nu}{}^{a b}$ is the Weyl conformal tensor. This action, leads to the well-know four-dimensional scale invariant Weyl action,
\begin{equation}
S =2a_{CG}\int \mathrm{d}^4 x\left(R_{\mu \nu} R^{\nu \mu}-\frac{1}{3} R^2\right).
\end{equation}

\section{Unification of Conformal Gravity with Internal \\Interactions, Fermions and Breakings}

\subsection{Unification and Field Content}

 In \cite{Roumelioti:2024lvn} was suggested that the unification of the CG with internal interactions  based on a framework that results in the GUT $SO(10)$ could be achieved using the $SO(2,16)$ as unifying gauge group. As it was emphasized in the Introduction the whole strategy was based on the observation that the dimension of the tangent space is not necessarily equal to the dimension of the corresponding curved manifold \cite{Roumelioti:2024jib,Percacci:1984ai,Percacci_1991,Nesti_2008,Nesti_2010,Krasnov:2017epi,Chamseddine2010,Chamseddine2016,noncomtomos,Konitopoulos:2023wst,Weinberg:1984ke}. An additional fundamental observation \cite{Roumelioti:2024lvn} is that in the case of $SO(2,16)$ one can impose Weyl and Majorana conditions
on fermions \cite{D_Auria_2001,majoranaspinors}. More specifically, using Euclidean signature for simplicity
(the implications of using non-compact space are explicitly
discussed in \cite{Roumelioti:2024lvn}), one starts with $SO(18)$ and with the fermions in its spinor representation, \textbf{256}. Then the spontaneous symmetry breaking of $SO(18)$ leads to its maximal subgroup $SO(6) \times SO(12)$ \cite{Roumelioti:2024lvn}. Let us recall for  convenience the branching rules of the relevant reps \cite{Li:1973mq,Slansky:1981yr,Feger:2019tvk},
\begin{align} 
     SO(18) &\supset SO(6) \times SO(12) \nonumber\\
   \textbf{256}   &= ( \textbf{4}, \overline{\textbf{32}}) + (\overline{\textbf{4}}, \textbf{32})\qquad \, \quad \!\qquad\qquad\qquad\qquad ~\text{spinor}\label{256spinor}\\
   \textbf{170}   &= (\textbf{1},\textbf{1}) + (\textbf{6},\textbf{12}) + (\textbf{20}',\textbf{1}) + (\textbf{1},\textbf{77})\qquad \quad \;\text{2nd rank symmetric}\label{170sym}
\end{align}
The breaking of $SO(18)$ to $SO(6) \times SO(12)$ is done by giving a vev to the $\langle \textbf{1},\textbf{1} \rangle$ component of a scalar in the $\textbf{170}$ rep. In turn, given that the Majorana condition can be imposed, due to the non-compactness of the used $SO(2,6) \sim SO(18)$, we are led after the spontaneous symmetry breaking to the $SO(6) \times SO(12)$ gauge theory with fermions in the $(\overline{\textbf{4}}, \textbf{32})$ representation.

 Then, according to \cite{Roumelioti:2024lvn}, the following spontaneous symmetry breakings can be achieved by using scalars in the appropriate representations.
\begin{equation}\label{cgtoeg}
 SO(6) \rightarrow SU(2) \times SU(2)~,        
\end{equation}
in the CG sector, and
\begin{equation}\label{so12toso10}
  SO(12) \rightarrow SO(10) \times [U(1)]_{global}
\end{equation}
in the internal gauge symmetry sector, with fermions in the $16_L (-1)$ under the $SO(10) \times [U(1)]_{global}$. The other generations are introduced as usual with more chiral fermions in the $256$ rep of $SO(18)$. In the present study, following \refse{AdjointSubsection},  we choose scalars in the 2nd rank antisymmetric $\textbf{15}$ rep of $SO(6)$ to break the CG gauge group, while the internal interactions gauge group $SO(12)$ is broken spontaneously by scalars in the $\textbf{77}$ rep. The $\textbf{15}$ rep can be drawn from the $SO(18)$ rep $\textbf{153}$:
\begin{equation}\label{153}
\textbf{153}   = (\textbf{15},\textbf{1}) + (\textbf{6},\textbf{12}) + (\textbf{1},\textbf{66})~,
\end{equation}
while from \refeq{170sym} we see that the $\textbf{77}$ rep can result from a $\textbf{170}$ rep of the parent group. Thus, in $SO(6)\times SO(12)$ notation, the scalars breaking the two gauge groups belong to $(\textbf{15},\textbf{1})$ and $(\textbf{1},\textbf{77})$, respectively.

According to the above picture we start from some high scale where the $SO(18)$ gauge group breaks, eventually  obtaining EG and $SO(10)\times[U(1)]_{global}$ after several  symmetry breakings. From that point, we use the symmetry breaking paths and field content followed in \cite{Djouadi:2022gws}, in order to finally arrive at the SM. In particular, the $SO(10)$ group breaks spontaneously into an intermediate group which eventually breaks into the SM gauge group. The intermediate groups are the Pati-Salam (PS) gauge group, $SU(4)_C\times SU(2)_L\times SU(2)_R$, with or without a discrete left-right symmetry, $\mathcal{D}$, and the minimal left-right gauge group (LR), $SU(3)_C\times SU(2)_L\times SU(2)_R\times U(1)_{B-L}$, again with or without the discrete left-right symmetry. We will denote the four intermediate gauge groups as 422, 422D, 3221 and 3221D, respectively.

The $SO(10)$ group breaks with a scalar $\textbf{210}$ into the 422 and the 3221D groups, with a scalar $\textbf{54}$ into 422 and with a scalar $\textbf{45}$ into 3221. The spontaneous breaking into the SM gauge group from each and every intermediate group is achieved with scalars that are accommodated in a 
$\overline{\textbf{126}}$ rep (still in $SO(10)$ language), while the Higgs boson necessary for the electroweak breaking will be accommodated in a $\textbf{10}$ rep\footnote{In \cite{Djouadi:2022gws} it is stated that, in order to accommodate the Higgs boson into a $\textbf{10}$ instead of a $\textbf{120}$ and to avoid an extra Yukawa term, a $U(1)$ Peccei-Quinn symmetry is taken into account. This could in principle be identified with the global $U(1)$ that survives the $SO(12)$ breaking and which also breaks at the unification scale.}. We will call the scale at which the $SO(10)$ gauge group breaks \textit{GUT scale}, $M_{GUT}$, in the sense that all three gauge couplings are unified at that scale, while we will call the scale at which the 422(D)/3221(D) groups break \textit{intermediate scale}, $M_I$. Thus, the consecutive breakings in each case can be seen below:

\footnotesize
\begin{align}
\text{422}: & \quad \text{SO(10)}|_{M_{GUT} } \xrightarrow{\langle \mathbf{210_H} \rangle} \, \, \, SU(4)_C\times SU(2)_R\times SU(2)_R|_{M_{I} }\xrightarrow{\langle \mathbf{\overline{126}_H}\rangle} \text{SM}\, ; \label{breakingchain1} \\
\text{422D}: & \quad \text{SO(10)}|_{M_{GUT} } \xrightarrow{\langle \mathbf{54_H} \rangle} ~~SU(4)_C\times SU(2)_R\times SU(2)_R \times {\cal D}|_{M_{I}}\xrightarrow{\langle \mathbf{\overline{126}_H}\rangle
} \text{SM}  \, ; \label{breakingchain2} \\
\text{3221}: & \quad \text{SO(10)}|_{M_{GUT}} \xrightarrow{\langle \mathbf{45_H} \rangle} \, \, \, SU(3)_C\times SU(2)_L\times SU(2)_R\times U(1)_{B-L}|_{M_{I}}\xrightarrow{\langle \mathbf{\overline{126}_H} \rangle} \text{SM}\, ; \label{breakingchain3}
\\
\text{3221D}: & \quad \text{SO(10)}|_{M_{GUT}} \xrightarrow{\langle \mathbf{210_H} \rangle} SU(3)_C\times SU(2)_L\times SU(2)_R\times U(1)_{B-L}\times {\cal D}|_{M_{I}}\xrightarrow{\langle \mathbf{\overline{126}_H} \rangle} \text{SM} \, .
\label{breakingchain4}
\end{align}
\normalsize
Considering the branching rules of: 
\begin{align} 
     SO(12) &\supset SO(10) \times [U(1)]_{global} \nonumber\\
   \textbf{12}   &= (\textbf{1})(2) + (\textbf{1})(-2) + ( \textbf{10})(0) \label{12}\\
   \textbf{66}   &= (\textbf{1})(0) + (\textbf{10})(2) +(\textbf{10})(-2) + ( \textbf{45})(0) \label{66}\\
   \textbf{77}   &= (\textbf{1})(4) +(\textbf{1})(0) +(\textbf{1})(-4) + (\textbf{10})(2) +(\textbf{10})(-2) + ( \textbf{54})(0) \label{77}\\
   \textbf{495}   &= (\textbf{45})(0)  + (\textbf{120})(2) +(\textbf{120})(-2) + ( \textbf{210})(0) \label{495}\\
   \textbf{792}   &= (\textbf{120})(0) +(\textbf{126})(0) +(\overline{\textbf{126}})(0) + (\textbf{210})(2) +(\textbf{210})(-2)\label{792}~,
\end{align}
we choose accommodate the Higgs $\textbf{10}$ rep into $\textbf{12}$ of $SO(12)$ and the $\overline{\textbf{126}}$ that breaks the intermediate gauge group into $\textbf{792}$. Regarding the four different breaking scenaria, $\textbf{45}$ will come from $\textbf{66}$, $\textbf{54}$ from $\textbf{77}$ and $\textbf{210}$ from $\textbf{792}$. Examining the $SO(18)$ following branching rules:
\begin{align} 
     SO(18) \supset& SO(6) \times SO(12) \nonumber\\
   \textbf{18}   =& ( \textbf{6}, \textbf{1}) + (\textbf{1}, \textbf{12})\label{18}\\
   \textbf{3060}   =& (\textbf{15},\textbf{1}) + (\textbf{10},\textbf{12}) + (\overline{\textbf{10}},\textbf{12}) + (\textbf{15},\textbf{66}) + (\textbf{6},\textbf{220}) + (\textbf{1},\textbf{495})\label{3060}\\
   \textbf{8568}   =& (\textbf{6},\textbf{1}) + (\textbf{15},\textbf{12})+ (\textbf{10},\textbf{66}) + (\overline{\textbf{10}},\textbf{66}) + (\textbf{15},\textbf{220}) + (\textbf{6},\textbf{495}) + \nonumber\\
   &+   (\textbf{1},\textbf{792})\label{8568}~,
\end{align}
and taking into account the branching rules of \refeqs{170sym} and (\ref{153}), we make the following choices regarding the $SO(12)$ reps: $\textbf{12}$ comes from $\textbf{18}$ of $SO(18)$, $\textbf{792}$ from $\textbf{8568}$, $\textbf{66}$ from $\textbf{153}$, $\textbf{495}$ from a $\textbf{3060}$ and $\textbf{77}$ from $\textbf{170}$.   For convenience, the accommodation of the full field content into the reps of each group is  given in \refta{content}.

\begin{center}
\begin{table}
\begin{center}
\renewcommand{\arraystretch}{2}
\begin{tabular}{|r|r|r|r|}
\hline
    $SO(10)$ & $SO(6)\times SO(12)$ & $SO(18)$  & Type  \& Role  \\\hline
 $\textbf{16}$ & $(\overline{\textbf{4}},\textbf{32})$ & $\textbf{256}$  & fermion, 3x generations   \\\hline
 - & $(\textbf{15},\textbf{1})$ &  $\textbf{153}$ & scalar, breaks $SO(6)$   \\\hline
 - & $(\textbf{1},\textbf{77})$ & $\textbf{170}$ & scalar, breaks $SO(12)$   \\\hline
 $\textbf{18}$ & $(\textbf{1},\textbf{12})$ & $\textbf{18}$$\textbf{18}$ & scalar, breaks SM   \\\hline
 $\overline{\textbf{126}}$ & $(\textbf{1},\textbf{792})$ & $\textbf{8568}$ & scalar, breaks the intermediate  groups into SM  \\\hline
 $\textbf{45}$ & $(\textbf{1},\textbf{66})$ & $\textbf{153}$ & scalar, breaks $SO(10)$ into 3221   \\\hline
 $\textbf{210}$ & $(\textbf{1},\textbf{495})$ & $\textbf{3060}$ & scalar, breaks $SO(10)$ into 422 \& 3221D   \\\hline
 $\textbf{54}$ & $(\textbf{1},\textbf{77})$ & $\textbf{170}$ & scalar, breaks $SO(10)$ into 422D   \\\hline
\end{tabular}
\caption{The full field content with the respective representations under each gauge group. 
}
\label{content}
\end{center}
\end{table}
\end{center}

\subsection{An estimation of the scales of spontaneous symmetry breakings}


With the the field content clear, we can now make an estimation of the scales at which all the above-mentioned breakings occur. This is achieved by the 1-loop running of the the gauge couplings in each energy regime.

\begin{figure}
\centering
\includegraphics[width=.4\textwidth]{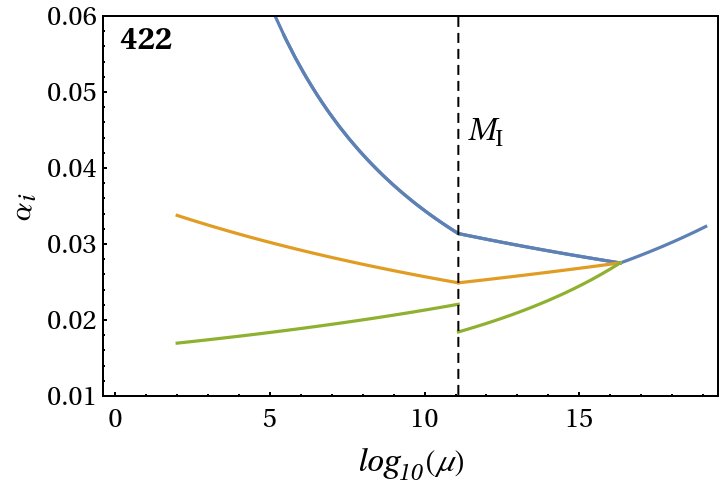}
\includegraphics[width=.4\textwidth]{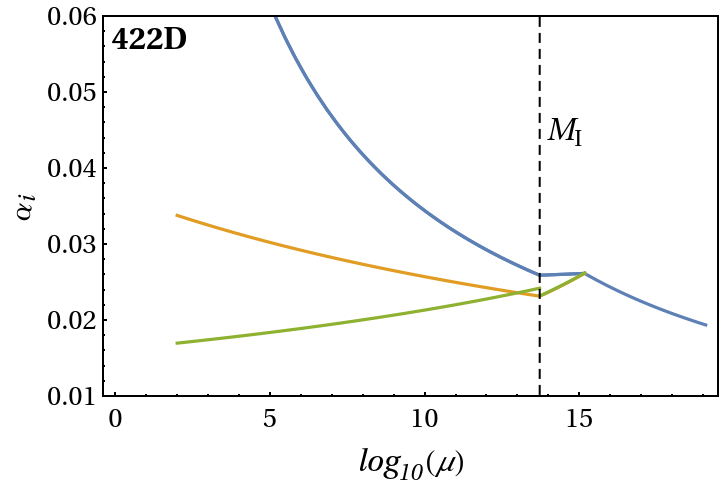}
\includegraphics[width=.4\textwidth]{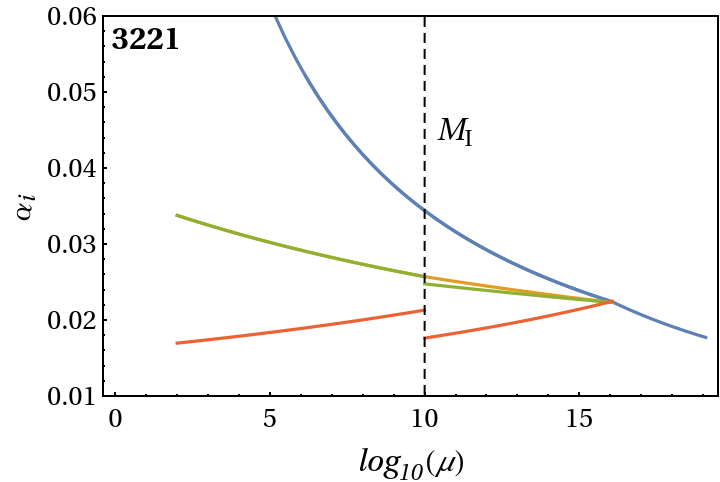}
\includegraphics[width=.4\textwidth]{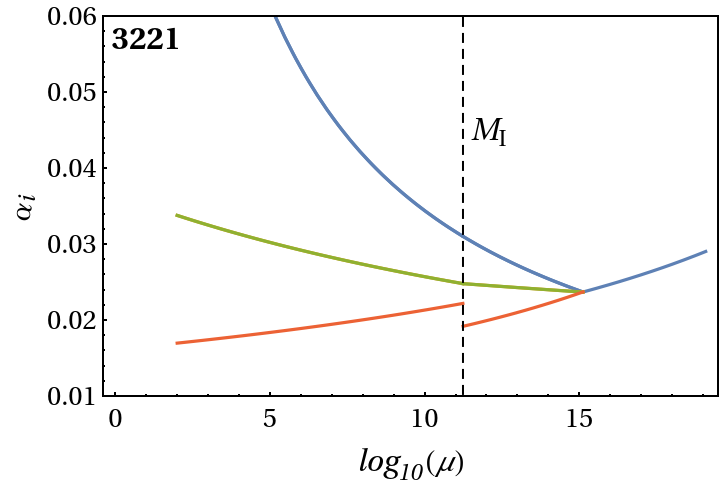}
\caption{The RG evolution of gauge couplings from the Electroweak scale to the limit close to the Planck scale is shown for the four cases. Top left: 422. Top right: 422D. Bottom left: 3221. Bottom right: 3221D.}
\label{low-ene}
\end{figure}

We begin from the $M_Z$ scale and the SM, where the values of all three gauge couplings are well known from the experiment  \cite{ParticleDataGroup:2022pth}. Then, using the 1-loop gauge $\beta$-functions for the SM energy regime and for each of the four intermediate gauge symmetries (which can be found in the Appendix, we find the intermediate scale $M_I$ and the GUT scale $M_{GUT}$ that allow for gauge coupling unification, and also the value of the unified gauge coupling at the GUT scale, $g_{10}(M_{GUT})$.  These results can be found in \refta{low-energy}. The matching conditions for the 422 breaking to the SM at the intermediate scale are:
\begin{align*}
\alpha_4^{422}(M_I)=&\alpha_3^{SM}(M_I)\\
\alpha_{2L}^{422}(M_I)=&\alpha_2^{SM}(M_I)\\
\frac{1}{\alpha_{2R}^{422}(M_I)}=&-\frac{2}{3}\frac{1}{\alpha_{3}^{SM}(M_I)}+\frac{5}{3}\frac{1}{\alpha_{1}^{SM}(M_I)}~,
\end{align*}
while the matching conditions for the 3221 breaking to the SM are:
\begin{align*}
\alpha_3^{3221}(M_I)=&\alpha_3^{SM}(M_I)\\
\alpha_{2L}^{3221}(M_I)=&\alpha_{2R}^{3221}(M_I)=\alpha_2^{SM}(M_I)\\
\frac{1}{\alpha_{1}^{3221}(M_I)}=&\frac{5}{2}\frac{1}{\alpha_{1}^{SM}(M_I)}-\frac{3}{2}\frac{1}{\alpha_{2}^{SM}(M_I)}~.
\end{align*}
Their RG evolution along the energy scale for each of the four cases is shown in \reffi{low-ene}.
\begin{center}
\begin{table}
\begin{center}
\renewcommand{\arraystretch}{2} 
\begin{tabular}{|l|r|r|r|}
\hline
       & $M_I$ (GeV) & $M_{GUT}$ (GeV) & $g_{10}^{(1)}(M_{GUT})$  \\\hline
 422   & $1.2\times10^{11}$ & $2.1\times10^{16}$  & $0.587$   \\\hline
 422D  & $5.2\times10^{13}$ & $1.5\times10^{15}$  & $0.572$   \\\hline
 3221  & $1.0\times10^{10}$ & $1.1\times10^{16}$  & $0.531$   \\\hline
 3221D & $1.7\times10^{11}$ & $1.4\times10^{15}$  & $0.546$   \\\hline
\end{tabular}
\caption{1-loop results for the intermediate scale, $M_I$, the GUT scale, $M_{GUT}$ and the unified gauge coupling at the GUT scale. 
}
\label{low-energy}
\end{center}
\end{table}
\end{center}
At this point we should note that the breaking of the CG to EG given in eq. \eqref{cgtoeg} gives a negative contribution to the cosmological constant and, if this was the only contribution, the space would be AdS. However, we have contributions to the cosmological constant from the other spontaneous breakings of the theory, i.e. the those of $SO(18)$ and $SO(12)$, which are positive. By choosing either of these breakings to be at the same scale with the breaking of the conformal gravity, the various contributions can be fine tuned to give a value for the cosmological constant of either zero or slightly positive, in agreement with the experimental observation.

\begin{figure}
\centering
\includegraphics[width=.8\textwidth]{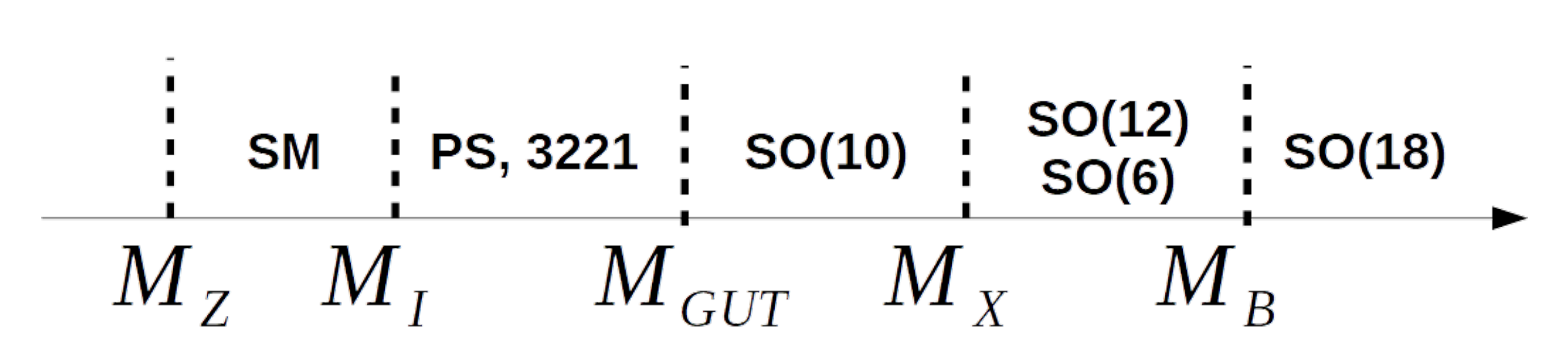}
\caption{The hierarchy among the various symmetry breaking scales of the models and the respective gauge groups that act on each energy regime.}
\label{scales}
\end{figure}

We will examine three different scenaria regarding the breakings above the GUT scale. In the first, named scenario \textit{A}, the $SO(18)$ gauge group breaks into $SO(6)\times SO(12)$ and they consequently break into EG and $SO(10)$ (and the global $U(1)$), all at the same scale,  $M_X$. This way, it is the contribution from the $SO(18)$ breaking to the cosmological constant that cancels the negative one that comes from the CG breaking. In contrast, in the scenaria  \textit{B} and \textit{C}, the $SO(18)$ gauge group breaks into $SO(6)\times SO(12)$ at a scale $M_B$, while both $SO(6)$ and $SO(12)$ will break at a different scale, $M_X$, which will be of course between $M_{GUT}$ and $M_B$.
The hierarchy among the scales is shown in \reffi{scales}.

It is important to note that, while for the evolution of the couplings after the breaking of $SO(10)$ the calculation is clear and we can determine the renormalization group equations (RGEs) in a straightforward way, when we turn to the running of gauge theories based on non-compact groups, the situation certainly is not clear. There exist serious calculations of the $\beta$-functions of the various terms of Stelle's $R^2$ gravity, which was proven to be renormalizable \cite{Stelle:1976gc,Stelle:1977ry}, but all calculations are done in Euclidean space \cite{Fradkin:1981iu,Avramidi:1985ki,Codello:2006in,Niedermaier:2009zz,Niedermaier:2010zz,Ohta:2013uca}. Therefore, strictly speaking the calculation of the $\beta$-function of a gauge theory based on a non-compact group has not been done. We speculate though that at least at one-loop level, the $\beta$-functions of gauge theories of non-compact groups could be well approximated by the corresponding ones of the compact ones. This speculation finds support from the suggestions of Donoghue in a number of papers \cite{Donoghue:2016xnh,Donoghue:2016vck,Donoghue:2017fvm}, which we adapt in the corresponding calculations of the $\beta$-functions presented in the Appendix.

Starting from scenario \textit{A}, if we run the $SO(10)$ gauge coupling until the $M_X$ scale, there it will match the value of the $SO(6)$ and $SO(12)$ gauge couplings, so
\begin{equation}
\alpha_{10}^{(1)}(M_X)=\alpha_{CG}^{(1)}(M_X)~. \label{mc}
\end{equation}
Considering the last term of \refeq{so24finalaction} and substituting the above relation, we can compare this term with the $SO(18)$ contributions to the cosmological constant. This way we can have an estimate of the breaking scale:
\begin{equation}
    M_X \sim  10^{18}~\text{GeV}~. \label{MxA}
\end{equation}
The precise value depends on various parameters, but the order of magnitude should remain as above. However, trying to run the $SO(18)$ coupling up to the Planck scale, we see that its steep RGE moves it almost immediately in the non-perturbative regime and it has a Landau pole lower than the Planck scale. As a consequence, for this scenario to work one would need to either drastically change the field content, or include additional new physics phenomena below the Planck scale.

Turning our attention to scenaria \textit{B} and \textit{C}, the  difference between them is that for scenario \textit{B}  we choose to break $SO(18)$ at a scale lower than the Planck scale, so $M_B<M_{Pl}$, while in scenario \textit{C} $SO(18)$ breaks exactly at the Planck scale, $M_B=M_{Pl}$. Then, in both scenaria, the $SO(6)\times SO(12)$ gauge group will run until the breaking scale $M_X$, under which we are left with the $SO(10)$ gauge group and EG. Note that $SO(6)$ and $SO(12)$ have to break at the same scale, if we want to keep the cosmological constant fine tuning.

Here  we do not have the  matching condition of \refeq{mc}, as we now have $\alpha_{10}^{(1)}(M_X)=\alpha_{12}^{(1)}(M_X)$. Thus,  using the $\beta$-function of $SO(10)$ and $SO(12)$ and the approximative $\beta$-functions of $SO(6)$, we make a rough estimate of the scale in question, which again is $M_X\sim10^{18}$GeV. For both scenaria, above $M_X$ the $SO(12)$ gauge coupling runs smoothly up to $M_B$, well within the perturbative regime. This is due to our choice of representations for some of the scalars of the theory (as explained above), since the fact that they are singlets under the CG gauge group avoids unwanted multiplicities during the calculation of the $SO(12)$ gauge $\beta$-function. For scenario  \textit{B}, $M_B$ is slightly below the Planck scale, and the $SO(18)$ gauge coupling runs up to $M_{Pl}$ without entering the non-perturbative regime, although its RG evolution is very steep. The runnings for scenaria \textit{B} and \textit{C} are can be found in \reffi{b} and \reffi{c}, respectively. Once more, all the $\beta$-functions can be found in the Appendix.

\begin{figure}
\centering
\includegraphics[width=.4\textwidth]{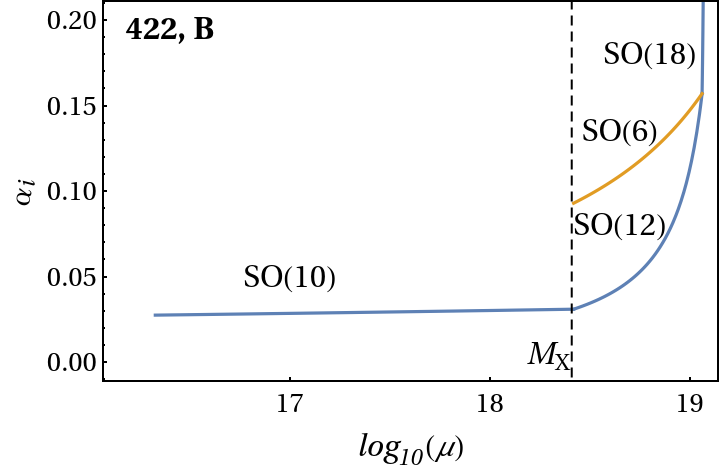}
\includegraphics[width=.4\textwidth]{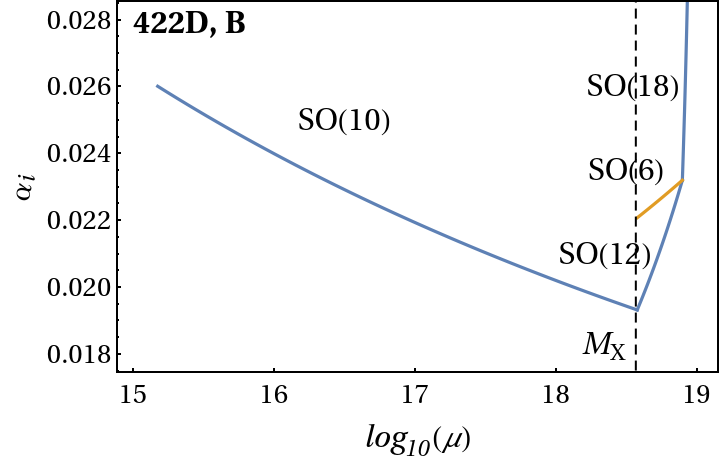}
\includegraphics[width=.4\textwidth]{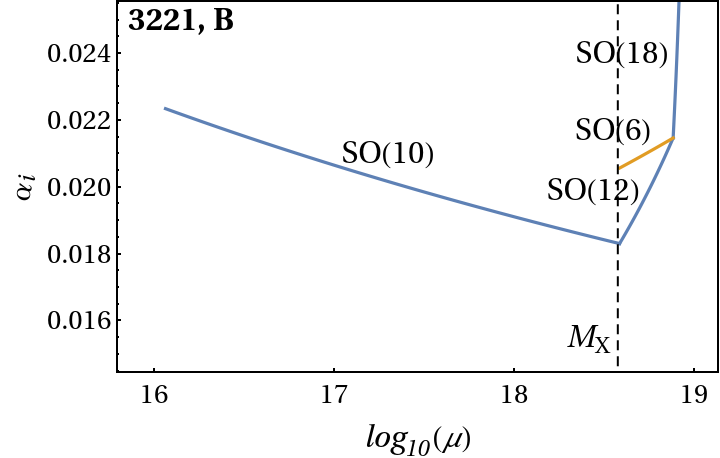}
\includegraphics[width=.4\textwidth]{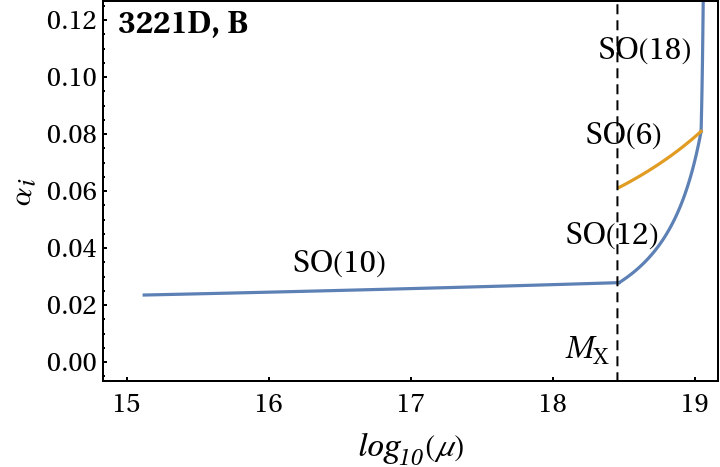}
\caption{The RG evolution of gauge couplings from the GUT scale up to the Planck scale is shown for the four cases for scenario \textit{B}. Top left: 422. Top right: 422D. Bottom left: 3221. Bottom right: 3221D.}
\label{b}
\end{figure}

\begin{figure}
\centering
\includegraphics[width=.4\textwidth]{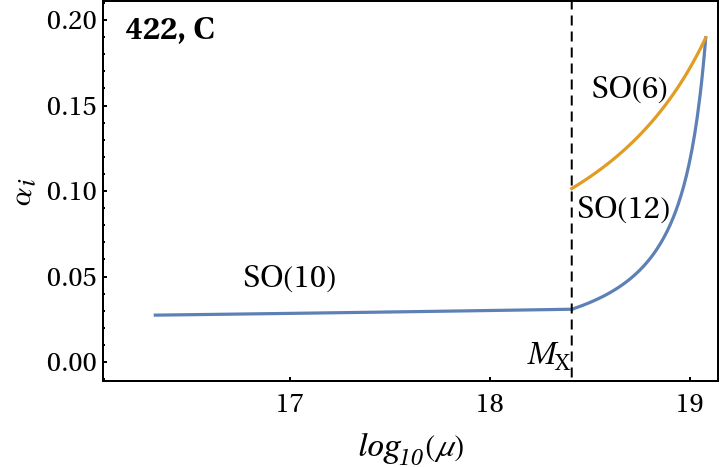}
\includegraphics[width=.4\textwidth]{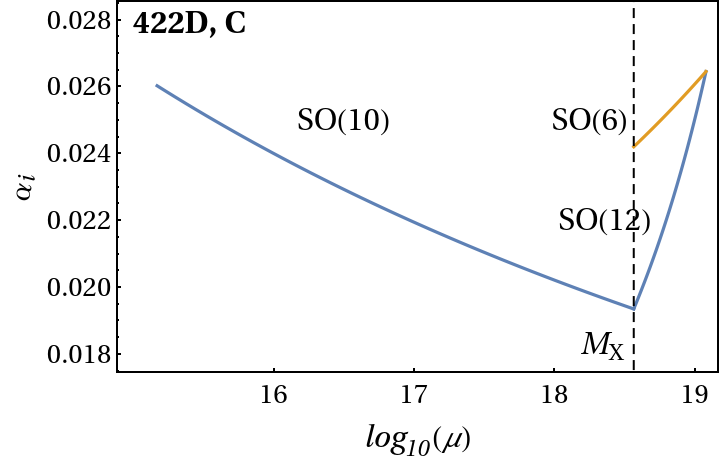}
\includegraphics[width=.4\textwidth]{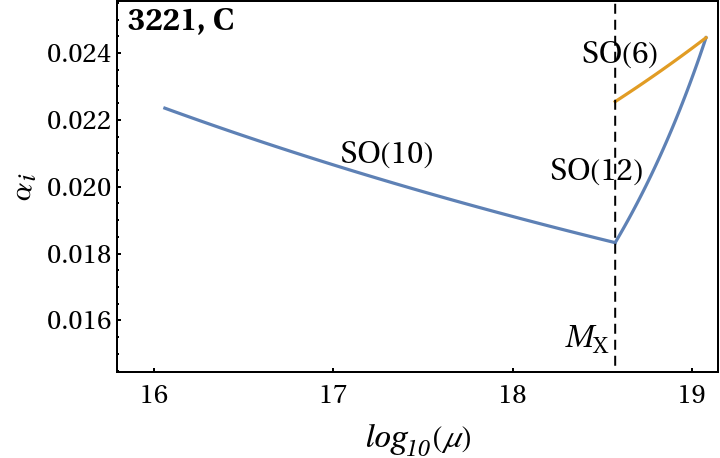}
\includegraphics[width=.4\textwidth]{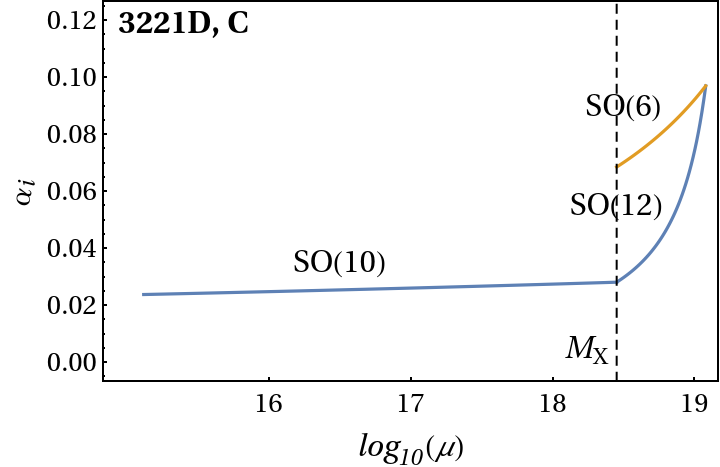}
\caption{The RG evolution of gauge couplings from the GUT scale up to the Planck scale is shown for the four cases for scenario \textit{C}. Top left: 422. Top right: 422D. Bottom left: 3221. Bottom right: 3221D.}
\label{c}
\end{figure}

We would like to add a comment about the case of FG. 
As it was explained in \cite{Roumelioti:2024jib}, when attempting to unify FG with internal interactions, along the lines of Unification of Conformal Gravity with $SO(10)$ \cite{Roumelioti:2024lvn},  the difficulties that in principle one is facing are that fermions should (a) be chiral in order to have a chance to survive in low energies and not receive masses as the Planck scale, (b) appear in a matrix representation, since the constructed FG is a matrix
model. Then it was suggested \cite{Roumelioti:2024jib} and given that the Majorana condition can be imposed, a solution satisfying the conditions (a) and (b) above is the following. We choose to start with the $SO(6) \times SO(12)$ as the initial
gauge theory with fermions in the $(\overline{\textbf{4}},\textbf{32})$ representation satisfying in this way the criteria to obtain chiral fermions in tensorial representation of a fuzzy space. Another important point is that using the gauge-theoretic formulation of gravity to construct the FG one is led to in gauging the $SO(6) \times U(1) \sim SO(2,4) \times U(1)$. Therefore from this point of view there exist only a small difference as compared to the CG.
Since the extra $U(1)$ symmetry of the gravity part is irrelevant to the above calculations, one could identify scenario \textit{C} to the  FG model. Obviously, the notable difference will be that there is no ($SO(18)$) group above the Planck scale, but this does not affect the calculation above.

\section{Conclusions and Discussion}
  In a previous paper \cite{Roumelioti:2024lvn}, a potentially realistic model was
constructed based on the idea that unification of gravity and internal
interactions in four dimensions can be achieved by gauging an enlarged
tangent Lorentz group. This possibility was based on the observation that
the dimension of the tangent space is not necessarily equal to the dimension of the corresponding curved manifold. In \cite{Roumelioti:2024lvn}, due to the very
interesting fact that gravitational theories can be described by gauge
theories, first was constructed the CG in a gauge theoretic
manner by gauging the $SO(2,4)$ group. Of particular interest was the fact
that the spontaneous symmetry breaking of the constructed CG could lead,
among others, to the EG and the WG. Then it is was possible to unify the
CG with internal interactions based on the $SO(10)$ GUT (after breaking of
$SO(12)$), using the higher-dimensional tangent group $SO(2,16)$. Inclusion of
fermions and application suitably the Weyl and Majorana conditions led to
a fully unified scheme, which was promised to be examined further concerning
its low scale behaviour as well as its cosmological predictions.

In the present work the behaviour of the fully unified theory has been
examined in some detail by examining its breakings at three scales, namely (i) the  scale that the $SO(18)$ gauge group breaks, $M_B$, (ii) the scale where $SO(6)\times SO(12)$ breaks, $M_X$, and (iii) the  $SO(10)$ unification scale of internal interactions, $M_{GUT}$ (the examination of the model includes the scale that the intermediate gauge groups below the GUT scale break into the SM gauge group, $M_I$, and it runs all the way down to the electroweak scale). 

This leads to three distinct scenaria. In scenario \textit{A}  we assume $M_B=M_X$, so $SO(18)$ breaks immediately to $SO(10)$ and EG, and CG and $SO(12)$ do not run. In scenario \textit{B} none of the above-mentioned scales coincide, so all different gauge structures run in their respective  energy regimes. In scenario \textit{C}, we identify $M_B=M_{Pl}$, so there is no $SO(18)$ running below the Panck scale.
The running of the usual internal, i.e. non-gravitational interactions,
has been also examined below the $SO(10)$ breaking scale. In particular, gauge couplings run down to the scale that the intermediate gauge groups break, $M_I$, and in turn the SM gauge couplings  are run all the way down to the electroweak scale.
The result is
that the proposed in \cite{Roumelioti:2024lvn} unification scheme of all interactions,
including gravity in the form of CG, which subsequently was broken to EG
is a realistic one and can be examined perturbatively for scenaria \textit{B} and \textit{C}, while we provide an estimate of all the symmetry breaking scales for each case. The FG case exhibits the same behaviour as scenario \textit{C}.

In the future we plan to study the cosmological implications of the
constructed Unified scheme. In particular we plan to do studies along
those done in the ghost-free bigravity \cite{Hassan:2013pca} and of E. Kolb and
collaborators \cite{Kolb:2023dzp}. A more immediate examination concerns the implications of the various spontaneous symmetry breakings discussed in the present paper in the formation of cosmic strings and their possible gravitational wave signal along the study of \cite{King:2021gmj}.

  It is always useful to repeat the reason that the present unified scheme
overcomes the Coleman–Mandula (CM) theorem \cite{Coleman1967}. The point being that
the CM theorem has several hypotheses and the most relevant is that
the theory is Poincaré invariant. In \cite{Roumelioti:2024lvn}, given that the
final aim was to obtain the Einstein gravitational theory coupled to
the GUTs, obviously the original conformal group, which is an extension of
the Poincaré group was spontaneously broken to the Lorentz group.

\section*{Acknowledgements}
It is a pleasure to thank Carmelo Martin and Tomas Ortin for discussions and Tom Kephart and Roberto Percacci for correspondence on the present work. Discussions on various stages of development of the theories discussed in the present work with our collaborators Thanassis Chatzistavrakidis, Alex Kehagias, Spyros Konitopoulos, George Manolakos, Pantelis Manousselis, Stelios Stefas, Manos Sadirakis and in particular with Danai Roumelioti are also very much appreciated.

GP is supported by the Portuguese Funda\c{c}\~{a}o para a Ci\^{e}ncia e Tecnologia (FCT) under Contracts UIDB/00777/2020, and UIDP/00777/2020, these projects are partially funded through POCTI (FEDER), COMPETE, QREN, and the EU. GP has a postdoctoral fellowship in the framework of UIDP/00777/2020 with reference BL154/2022\_IST\_ID. GZ would like to thank IFT-Madrid,
MPP-Munich, ITP-Heidelberg, and DFG Exzellenzcluster 2181:STRUCTURES of Heidelberg University for their hospitality and support.

\appendix
\section*{Appendix:~$\beta$-Functions of the Gauge Couplings at each Energy Sale}\label{AppA}

In this section we give all the one-loop $\beta$-functions of the gauge couplings of all gauge groups -with the appropriate field content each time-  that we encounter after the various spontaneous symmetry breakings mentioned above. They are given by:
\begin{align}
    \beta_{g_i}=16\pi^2\mu\frac{d}{d\mu}g_i=b_ig_i^3 \quad \quad \text{or}\quad\quad
    \beta_{\alpha_i}=2\pi\mu\frac{d}{d\mu}\alpha_i=b_i\alpha_i^2~,
\end{align}
where $\alpha_i=g_i^2/4\pi$ and $b_i$ is the respective $\beta$-function coefficient, which is the quantity we need to calculate in each case.

Considering the group-theoretical details of all three SM gauge groups well-known, we can give their respective $\beta$-function coefficients:
\begin{align}
    b_3=&-7\\
    b_2=&-\frac{19}{6}\\
    b_1=&\frac{41}{10}~,
\end{align}
where the $b_1$ ocoefficient is given in the usual $SU(5)$ normalization. \\

\noindent For the four intermediate gauge groups using the usual way of \cite{Jones:1981we}, the $b_i$ coefficients can be easily found to be:

\vspace{0.2cm}

\begin{multicols}{4}

\underline{442}:\\

\begin{itemize}
\item $b_4=-\frac{7}{3}$
\item $b_{2L}=2$
\item $b_{2R}=\frac{28}{3}$
\end{itemize}

\phantom{a}

\columnbreak

\underline{442D}:\\

\begin{itemize}
\item $b_4=\frac{2}{3}$
\item $b_{2L}=\frac{28}{3}$
\item $b_{2R}=\frac{28}{3}$
\end{itemize}

\phantom{a}

\columnbreak

\underline{3221}:\\

\begin{itemize}
\item $b_3=-7$
\item $b_{2L}=-\frac{8}{3}$
\item $b_{2R}=-2$
\item $b_1=\frac{11}{2}$
\end{itemize}

\columnbreak

\underline{3221D}:\\

\begin{itemize}
\item $b_3=-7$
\item $b_{2L}=-\frac{4}{3}$
\item $b_{2R}=-\frac{4}{3}$
\item $b_1=7$
\end{itemize}

\end{multicols}

\vspace{0.2cm}

\noindent For the bigger groups that feature in the above theory, we give a list of a few details that are necessary for the calculation of their $b_i$'s:

\vspace{0.2cm}

\begin{multicols}{4}

\underline{$SO(6)$}:\\

\begin{itemize}
\item Generators:\\ $N^2-1=15$
\item $C_2(\textbf{15})=4$
\item $T(\textbf{15})=4$
\item $T(\textbf{4})=1/2$
\end{itemize}

\phantom{a}\\
\phantom{a}\\
\phantom{a}\\
\phantom{a}\\
\phantom{a}\\
\phantom{a}\\

\columnbreak

\underline{$SO(10)$}:\\

\begin{itemize}
\item Generators:\\ $N(N-1)/2=45$
\item $C_2(\textbf{45})=$\\$T(\textbf{45})=8$
\item $T(\textbf{16})=2$
\item $T(\textbf{10})=1$
\item $T(\textbf{54})=12$
\item $T(\textbf{126})=35$
\item $T(\textbf{210})=56$
\end{itemize}

\phantom{a}

\columnbreak

\underline{$SO(12)$}:\\

\begin{itemize}
\item Generators:\\ $N(N-1)/2=66$
\item $C_2(\textbf{66})=$\\$T(\textbf{66})=10$
\item $T(\textbf{12})=1$
\item $T(\textbf{32})=4$
\item $T(\textbf{77})=14$
\item $T(\textbf{495})=120$
\item $T(\textbf{792})=210$
\end{itemize}

\phantom{a}\\

\columnbreak

\underline{$SO(18)$}:\\
\begin{itemize}
\item  Generators:\\ $N(N-1)/2=153$
\item $C_2(\textbf{153})=$\\$T(\textbf{153})=16$
\item  $T(\textbf{18})=1$
\item $T(\textbf{170})=20$
\item $T(\textbf{256})=32$
\item $T(\textbf{3060})=$\\$560$
\item $T(\textbf{8568})=$\\$1820$
\end{itemize}

\end{multicols}


\noindent With this information at hand, we can now calculate their respective $b_i$ coefficients for each intermediate breaking scenario:

\vspace{0.2cm}

\begin{multicols}{4}

\underline{$SO(10)$}:\\

\begin{itemize}
\item $b_{10}^{422}=\frac{16}{3}$
\item $b_{10}^{422D}=-\frac{28}{3}$
\item $b_{10}^{3221}=-\frac{32}{3}$
\item $b_{10}^{3221D}=\frac{16}{3}$
\end{itemize}

\columnbreak

\underline{$SO(12)$}:\\

\begin{itemize}
\item $b_{12}^{422}=\frac{331}{3}$
\item $b_{12}^{422D}=75$
\item $b_{12}^{3221}=\frac{221}{3}$
\item $b_{12}^{3221D}=\frac{331}{3}$
\end{itemize}

\columnbreak

\underline{$SO(6)$}:\\

\begin{itemize}
\item $b_{6}^{422}=4$
\item $b_{6}^{422D}=4$
\item $b_{6}^{3221}=4$
\item $b_{6}^{3221D}=4$
\end{itemize}

\columnbreak

\underline{$SO(18)$}:\\

\begin{itemize}
\item $b_{18}^{422}=-$
\item $b_{18}^{422D}=-$
\item $b_{18}^{3221}=-$
\item $b_{18}^{3221D}=-$
\end{itemize}

\end{multicols}

\vspace{0.2cm}

\noindent One can see that, since all the scalars that break the intermediate symmetries at $M_I$ are singlets under the CG group, its $b_i$'s will be the same for each case.



\bibliographystyle{h-physrev5}
\bibliography{main}

\end{document}